\documentstyle[prl,aps,epsf,floats,color]{revtex}
\begin{document}
\baselineskip=12pt
\def\be{\begin{equation}}
\def\ee{\end{equation}}
\def\bea{\begin{eqnarray}}
\def\eea{\end{eqnarray}}
\def\E{{\rm e}}
\def\bearst{\begin{eqnarray*}}
\def\eearst{\end{eqnarray*}}
\def\peleven{\parbox{11cm}}
\def\peffec{\peight{\bearst\eearst}\hfill\peleven}
\def\pspace{\peight{\bearst\eearst}\hfill}
\def\ptwelve{\parbox{12cm}}
\def\peight{\parbox{8mm}}
\twocolumn[\hsize\textwidth\columnwidth\hsize\csname@twocolumnfalse\endcsname

\title
{Long range correlation in cosmic microwave background radiation}

\author{M. Sadegh Movahed$^{1,2}$, F. Ghasemi$^3$, Sohrab Rahvar$^{4}$, M. Reza Rahimi Tabar$^{4,5}$}
\address{$^{1}$ Department of Physics, Shahid Beheshti University, G.C., Evin, Tehran 19839, Iran}
\address{$^{2}$ School of Astronomy, Institute for Research in Fundamental Sciences, (IPM), P.O. Box 19395-5531,
Tehran, Iran}
\address{$^{3}$ Department of Chemical Engineering \& Materials Science, University of
Minnesota, Minneapolis, Minnesota 55455}
\address{$^{4}$Department of Physics, Sharif University of
Technology, P.O.Box 11365--9161, Tehran, Iran}
\address{$^{5}$Fachbereich Physik, Universit\"{a}t Osnabr\"{u}ck, Barbarastra{\ss}e
7, 49076 Osnabr\"{u}ck, Germany}

\vskip 1cm

 \maketitle
%\date{00/07/2000}
%\maketitle

%%%%%%%%%%%%%%%%%%%%%%%%%%%%%%%%%%%%%%%%%%%%%%%%%%%%%%
%ABSTRACT
%%%%%%%%%%%%%%%%%%%%%%%%%%%%%%%%%%%%%%%%%%%%%%%%%%%%%%

\begin{abstract}

 We investigate the statistical anisotropy and Gaussianity of temperature
fluctuations of Cosmic Microwave Background radiation (CMB) data
from {\it Wilkinson Microwave Anisotropy Probe} survey, using the
multifractal detrended fluctuation analysis, rescaled range and
scaled windowed variance methods. The multifractal detrended
fluctuation analysis shows that CMB fluctuations has a long range
correlation function with a multifractal behavior. By comparing the
shuffled and surrogate series of CMB data, we conclude that the
multifractality nature of temperature fluctuation of CMB is mainly
due to the long-range correlations and the map is
consistent with a Gaussian distribution.
\end{abstract}
\hspace{.3in}
\newpage
]

\section{Introduction}

The {\it Wilkinson Microwave Anisotropy Probe} ({\it WMAP}) mission
is designed to determine the geometry, matter content, and the
evolution of the universe. It is shown that the universe is
geometrically flat and dark energy at the present time has the
dominant contribution in the matter content of the universe, causes
the universe to have positive acceleration
\cite{af04,ben03,spe03,teg04}. The statistical properties of the
Cosmic Microwave Background radiation (CMB) data can be a unique
tool to identify the parameters of standard model of cosmology
\cite{pei03}. One of the main aims of the statistical analysis of
the CMB is to examine the early universe scenarios as the
inflationary cosmology. The Gaussianity of primordial fluctuations
is a key assumption of modern cosmology, motivated by inflationary
models \cite{mu81,haw82,guth85}.

Since the observed CMB sky is a single realization of the map of the
large scale structures, the detection of statistical isotropy
violation or correlation patterns pose a great observational
challenge. In order to extract more information from the rich source
of information provided by present (and future) CMB maps
\cite{planck1,planck2,spt}, it is important to design as many
independent statistical methods as possible to study deviations from
standard statistics, such as statistical isotropy and possible
correlations. Since statistical isotropy can be violated in many
different ways \cite{ack07}, various statistical methods can come to
different conclusions. Each method by design is more sensitive to a
special kind of statistical isotropy violation
\cite{zhen10,amir10,tarun10}. Based on bipolar power spectrum, it
has shown no strong evidence of statistical isotropy violation
\cite{hajian05,hajian04,amir06}, but analysis of the distribution of
extrema in {\it WMAP} sky maps has indicated non-Gaussianity and to
some extent, violation of statistical isotropy \cite{larson}. In
addition there are many criteria have been introduced to measure the
statistical isotropy of CMB such as, quadratic maximum-likelihood
\cite{dun09}, multipole vectors \cite{car10} and more recently P.
Naselsky et. al. used symmetry of CMB map to determine statistical
isotropy as well as non-Gaussianity \cite{nas11}.

The statistical properties of the primordial fluctuations generated
by the inflationary cosmology are closely related to the cosmic
microwave background radiation anisotropy and a measurement of
non-Gaussianity is a direct test of the inflation paradigm (see Refs. \cite{barto10,rau10} for more details).
% If CMB anisotropy is Gaussian, then the angular power spectrum fully
% specifies its statistical properties.
Many authors have searched for non-Gaussian signatures in CMB data
using peak distributions \cite{bar86,bond87}, the genus curve
\cite{col88,smoot94}, peak correlations \cite{kog96}, global
Minkowski functionals methods \cite{win97} and directional spherical
wavelet \cite{mec08,mec081}. The techniques used for the detection
of non-Gaussianity in hydrodynamic turbulence was applied for CMB
data \cite{sree03,sa05a}. Moreover the Gaussianity of CMB in
different angular scales have been tested
\cite{heavens98,peak85,alan99,alan01,schmalzing98,ferreira98,pando98,bromley99,banday00,contaldi00,mukherjee00,magueijo00,novikov00,sandvik01,barreiro00,phillips01,komatsu02,komatsu01,kunz01,aghanim01,cayon02,park01,shandarin02,wu01,santos02,polenta02,Mead2010,lew08,reath10,drag10}.
Most of the previous works only tested the consistency between the
CMB data and simulated Gaussian realizations and so far they found
no significant evidence for cosmological non-Gaussianity.

In this work we characterize the complex behavior of CMB through the
computation of the fluctuation parameters - scaling exponents -
which quantifies the correlation exponents and multifractality of
the data. We use certain fractal analysis approaches such as,
Multifractal Detrended Fluctuation Analysis (MF-DFA), Rescaled Range
Analysis $(R/S)$ and Scaled Windowed Variance (SWV) to analysis the
data set. Using the new approaches, we will test the statistical
isotropy and Gaussianity of temperature fluctuations at the last
scattering surface. The MF-DFA method shows that Cosmic Microwave
Background fluctuations has a long range correlation function with
multifractal behavior. Comparing the MF-DFA results of the original
temperature fluctuations to those for shuffled and surrogate series,
we conclude that the multifractality nature of CMB is mainly due to
long-range correlations and the map is consistent with Gaussian
distribution. Applied methods of MF-DFA, $R/S$ and SWV, show that
{\it WMAP} data is a statistically isotropic data set. The value of
scaling exponent (Hurst exponent) guarantees that there is no
evidence for violation of statistical isotropy in the CMB anisotropy
map considered here.

 This paper is organized as follows:  In Section II we briefly describe
  the MF-DFA method and show that the scaling exponents determined via the
MF-DFA method are identical to those obtained by the standard
multifractal formalism based on partition functions. Also briefly
introduce the Rescaled Range Analysis and Scaled Windowed Variance
methods. The fractal analysis of the temperature fluctuations are
presented in Section III. In Section IV, we compare the multifractal
behavior of the original data with that of shuffled and surrogate
series and show that the multifractality is mainly due to the
long-range correlations. In Section V we investigate the statistical
isotropy and Gaussianity of temperature fluctuations using the
MF-DFA, {\it R/S} and SWV results. Section VI provides
discussion and presents the results.\\

\section{Fractal Analysis Methods}

In this section we review three standard methods, i.e. MF-DFA, $R/S$
and SWV, to investigate the fractal properties of stochastic
processes.

\subsection{Multifractal Detrended Fluctuation Analysis}

The MF-DFA methods are the modified version of detrended fluctuation
analysis (DFA) to detect multifractal properties of series. The
detrended fluctuation analysis (DFA) method introduced by Peng et
al. \cite{Peng94} has became a widely-used technique for
determination of (mono)-fractal scaling properties and the detection
of long-range correlations in noisy non-stationary time series
\cite{Peng94,taqqu,physa,kunhu,kunhu1}. It has successfully been
applied to diverse fields such as DNA sequences \cite{Peng94,dns},
heart rate dynamics \cite{herz,PRL00}, neuron spiking \cite{neuron},
human gait \cite{gait}, long-time weather records \cite{wetter},
cloud structure \cite{cloud}, geology
\cite{malamudjstatlaninfer1999}, ethnology \cite{Alados2000},
economical time series \cite{economics}, solid state physics
\cite{fest}, solar physics and plasma fluctuations  \cite{sa05b}.

The simplest type of the multifractal analysis is based upon the
standard partition function multifractal formalism, which has been
developed for the multifractal characterization of normalized,
isotropic (stationary) measurements
\cite{feder88,barabasi,peitgen,bacry01}. Unfortunately, this
standard formalism does not give correct results for non-isotropic
angular (non-stationary time) series that are affected by trends
or that cannot be normalized. Thus, in the early 1990s an improved
multifractal formalism has been developed, the wavelet transform
modulus maxima (WTMM) method \cite{wtmm}, which is based on the
wavelet analysis and involves tracing the maxima lines in the
continuous wavelet transform over all scales. The other method,
the multifractal detrended fluctuation analysis (MF-DFA), is based
on the identification of scaling of the $q$th-order moments
depending on the signal length and is a generalization of the
standard DFA using only the second moment $q=2$.

The MF-DFA does not require the modulus maxima procedure in contrast
to the WTMM method, and hence does not require more effort in
programming and computing than the conventional DFA. On the other
hand, often experimental data are affected by non-isotropic
(non-stationarities) like trends, which have to be well
distinguished from the intrinsic fluctuations of the system in order
to find the correct scaling behavior of the fluctuations. In
addition very often we do not know the reasons for underlying trends
in collected data and even worse we do not know the scales of the
underlying trends, also, usually the available recorded data is
small. For the reliable detection of correlations, it is essential
to distinguish trends from the intrinsic fluctuations in the data.
Non-detrending methods work well if the records are long and do not
involve trends. But if trends are present in the data, they might
give wrong results. Detrended fluctuation analysis (DFA) is a
well-established method for determining the scaling behavior of
noisy data in the presence of trends without knowing their origin
and shape \cite{Peng94,dns,herz,fano,allan}

The modified multifractal DFA (MF-DFA) procedure consists of five
steps. The first three steps are essentially identical to the
conventional DFA procedure (see e.~g.
\cite{Peng94,taqqu,physa,kunhu,kunhu1}). In our case which is
studying the temperature fluctuations of CMB, we take the
temperature data  with the size of $N$ and follow the steps as
follows:

 \noindent $\bullet$ {\it Step 1}: Determine the ``profile''
\begin{equation} Y(\gamma_s) \equiv \sum_{i=1}^{s} \left[ T(\hat{n}_i) - \langle
T \rangle \right],\qquad s=1,\ldots,N,  \label{profile}
\end{equation}
where $ T(\hat{n}_i)$ is the temperature of CMB map, $\hat{n_i}$ is
the unit vector pointing CMB and $\gamma_s=\arccos(\hat{n}_1.
\hat{n}_s)$ is the size of segment of CMB that we are calculating
the series. Subtraction of the mean $\langle T \rangle$ is not
compulsory, since it would be eliminated by the later detrending in
the third step.

\noindent $\bullet$ {\it Step 2}: Divide the profile $Y(\gamma_s)$
into $N_{\gamma_s} \equiv {\rm int}(N/s)$ non-overlapping segments
of equal angular lengths $\gamma_s$.

\noindent $\bullet$ {\it Step 3}: Calculate the local trend for each
of the $N_{\gamma_s}$ segments by fitting a polynomial function to
$Y(\gamma_s)$. The variance between $Y(\gamma_s)$ and the function
of the best fit for each segment $\nu$, $\nu = 1, \ldots,
N_{\gamma_s}$, is as follows:
\begin{equation} F^2(\gamma_s,\nu) \equiv {1 \over s} \sum_{i=1}^{s}
\left\{ Y[(\nu-1) \gamma_s + \gamma_i] - y_{\nu}(\gamma_i)
\right\}^2, \label{fsdef}
\end{equation}

A Linear, quadratic, cubic, or higher order polynomials can be used
in the fitting procedure (conventionally called DFA1, DFA2, DFA3,
$\ldots$) \cite{Peng94,physa,kunhu,PRL00}.

\noindent $\bullet$ {\it Step 4}: Average over all segments to
obtain the $q$-th order fluctuation function, defined as:
\begin{equation} F_q(\gamma_s) \equiv \left\{ {1 \over N_{\gamma_s}}
\sum_{\nu=1}^{ N_{\gamma_s}} \left[ F^2(\gamma_s,\nu)
\right]^{q/2} \right\}^{1/q}, \label{fdef}\end{equation}
where, in general, the index variable $q$ can take any real value
except zero. For $q=0$, equation (\ref{fdef}) becomes:
\begin{equation}
F_0(\gamma_s)= \exp\left( {1 \over  2N_{\gamma_s}} \sum_{\nu=1}^{
N_{\gamma_s}}\ln F^2(\gamma_s,\nu)\right ) \label{fdef0}
\end{equation}
For $q=2$, the standard DFA is retrieved. Generally we are
interested in how the generalized $q$ dependent fluctuation
functions, $F_q(\gamma_s)$, depend on the angular scale $\gamma_s$
for different values of $q$. Hence, we must repeat steps 2, 3 and 4
for several angular scales $\gamma_s$.  It is apparent that
$F_q(\gamma_s)$ will increase with the increasing of $\gamma_s$.

\noindent $\bullet$ {\it Step 5}: Determine the scaling behavior
of the fluctuation functions by analyzing log-log plots of
$F_q(\gamma_s)$ versus $\gamma_s$ for each value of $q$. If the
series $T(\hat{n}_i)$ are long-range power-law correlated,
$F_q(\gamma_s)$ increases, for large values of $\gamma_s$, as a
power-law
\begin{equation} F_q(\gamma_s) \sim \gamma_s^{h(q)} \label{Hq}. \end{equation}
In general, the exponent $h(q)$ may depend on $q$. For isotropic
fluctuations, $0<h(2)<1.0$ and $h(2)$ is identical to the
well-known Hurst exponent $(h(2)=H)$ \cite{Peng94,taqqu,feder88}.
In the absence of statistical isotropy the corresponding scaling
exponent of $F_q(\gamma_s)$ is larger than unity $h(2)>1.0$ and
its relation to the Hurst exponent is $H=h(q=2)-1$
\cite{Peng94,sa05b,eke02}. Thus, one can call the function $h(q)$
as the generalized Hurst exponent.

For monofractal fluctuations, $h(q)$ is independent of $q$, since
the scaling behavior of the variances $F^2(\gamma_s,\nu)$ is
identical for all segments of $\nu$, and the averaging procedure
in Eq.~(\ref{fdef}) will just give this identical scaling behavior
for all values of $q$. If we consider positive values of $q$, the
segments $\nu$ with large variance $F^2(\gamma_s,\nu)$ (i.~e.
large deviations from the corresponding fit) will dominate the
average $F_q(\gamma_s)$.  Thus, for positive values of $q$, $h(q)$
describes the scaling behavior of the segments with large
fluctuations. Usually the large fluctuations are characterized by
a smaller scaling exponent $h(q)$ for multifractal series
\cite{bun02}. On the contrary, for negative values of $q$, the
segments $\nu$ with small variance $F^2(\gamma_s,\nu)$ will
dominate the average $F_q(\gamma_s)$. Hence, for negative values
of $q$, $h(q)$ describes the scaling behavior of the segments with
small fluctuations \cite{bun02}.

\subsubsection{ Relation to standard multifractal analysis}

For an isotropic series the multifractal scaling exponents $h(q)$
defined in Eq.~(\ref{Hq}) are directly related to the scaling
exponents $\tau(q)$ defined by the standard partition function-based
multifractal formalism as shown below. Suppose that the data
$T(\hat{n}_i)$ of length $N$ is an isotropic sequence. Then the
detrending procedure in step 3 of the MF-DFA method is not required,
since no trend has to be eliminated. Thus, the DFA can be replaced
by the standard Fluctuation Analysis (FA) with the definition of
variance for each segment $\nu$, $\nu = 1, \ldots, N_{\gamma_s}$ as
follows:

\begin{equation} F_{\rm FA}^2(\gamma_s,\nu) \equiv [Y(\nu \gamma_s) - Y((\nu-1) \gamma_s)]^2.
\label{FAfsdef} \end{equation}

Inserting this simplified definition into Eq.~(\ref{fdef}) and using
Eq.~(\ref{Hq}), we obtain:

\begin{equation} \left\{ {1 \over  N_{\gamma_s}} \sum_{\nu=1}^{ N_{\gamma_s}}
\vert Y(\nu \gamma_s) - Y((\nu-1) \gamma_s) \vert^q \right\}^{1/q}
\sim \gamma_s^{h(q)}. \label{FAfHq} \end{equation}

In order to relate also to the standard textbook box counting
formalism \cite{feder88,barabasi,peitgen,bacry01}, we employ the
definition of the profile in Eq.~(\ref{profile}). It is evident that
the term $Y(\nu \gamma_s) - Y((\nu-1) \gamma_s)$ in
Eq.~(\ref{FAfHq}) is identical to the sum of the numbers
$T(\hat{n}_i)$ within each segment $\nu$ of size $\gamma_s$. This
sum is known as the box probability $p_{\gamma_s}(\nu)$ in the
standard multifractal formalism for $T(\hat{n}_i)$:

\begin{equation} p_{\gamma_s}(\nu) \equiv \sum_{i=(\nu-1)s +1}^{\nu s} T(n_i) =
Y(\nu \gamma_s) - Y((\nu-1) \gamma_s).  \label{boxprob}
\end{equation}

The scaling exponent $\tau(q)$ is usually defined via the
partition function $Z_q(\gamma_s)$

\begin{equation} Z_q(\gamma_s) \equiv \sum_{\nu=1}^{N_{\gamma_s}} \vert p_{\gamma_s}(\nu)
\vert^q \sim \gamma_s^{\tau(q)}, \label{Zq} \end{equation}
where $q$
is a real parameter as in the MF-DFA method, discussed above. Using
Eq.~(\ref{boxprob}), we see that Eq.~(\ref{Zq}) is identical to
Eq.~(\ref{FAfHq}), and obtain analytically the relation between the
two sets of multifractal scaling exponents

\begin{equation} \tau(q) = q h(q) - 1. \label{tauH} \end{equation}
Thus, we observe that $h(q)$ defined in Eq.~(\ref{Hq}) for the
MF-DFA is directly related to the classical multifractal scaling
exponents $\tau(q)$.  Note that $h(q)$ is different from the
generalized multifractal dimensions which is defined as:

\begin{equation} D(q) \equiv {\tau(q) \over q-1}, \label{Dq} \end{equation}
that are used instead of $\tau(q)$ in some papers.  While $h(q)$ is
independent of $q$ for a monofractal time series, $D(q)$ depends on
$q$ in this case. Another way to characterize a multifractal series
is the singularity spectrum $f(\alpha)$, that is related to
$\tau(q)$ via a Legendre transform \cite{feder88,peitgen}
\begin{equation} \alpha = \tau'(q) \quad {\rm and} \quad
f(\alpha) = q \alpha - \tau(q). \label{Legendre} \end{equation}
here, $\alpha$ is the singularity strength or H\"older exponent,
while $f(\alpha)$ denotes the dimension of the subset of the
series that is characterized by $\alpha$. Using Eq.~(\ref{tauH}),
we can directly relate $\alpha$ and  $f(\alpha)$ to $h(q)$
\begin{equation} \alpha = h(q) + q h'(q) \quad {\rm and} \quad
f(\alpha) = q [\alpha - h(q)] + 1.\label{Legendre2} \end{equation} A
H\"older exponent denotes mono-fractality, while in the multifractal
case, the different parts of the structure are characterized by
different values of $\alpha$, leading to the existence of the
spectrum $f(\alpha)$. The interval of H$\ddot{\rm o}$lder spectrum
for a multifractal process, $\alpha\in [\alpha_{ min},\alpha_{
max}]$, can be determined by \cite{muzy94}
\begin{eqnarray}
\alpha_{ min}&=&\lim_{q \rightarrow +\infty} \frac{\partial \tau(q)}{\partial q}\\
\alpha_{max}&=&\lim_{q \rightarrow -\infty} \frac{\partial
\tau(q)}{\partial q}
\end{eqnarray}

\subsection{Scaled Windowed Variance Analysis}

The Scaled Windowed Variance analysis was developed by Cannon et al.
(1997)\cite{eke02}. The profile of temperature, $Y(\gamma_s)$, is
divided into intervals of angular length scale $\gamma_s$. Then the
standard deviation is calculated within each interval using the
following relation
\begin{equation}
{\rm
SWV}(\gamma_s)=\left(\frac{1}{s}\sum_{i=1}^{s}[Y(\gamma_i)-\langle
Y(\gamma_s)\rangle]^2\right)^{1/2}.
\end{equation}
  The average standard deviation of all angular intervals of length $\gamma_s$ is computed. This
computation is repeated over all possible interval lengths. The
scaled windowed variance is related to $\gamma_s$ by a power law
\begin{equation}\label{swv}
{\rm SWV}\sim \gamma_s^H.
\end{equation}

\subsection{Rescaled Range Analysis}

Hurst developed the Rescaled Range analysis, a statistical method to
analyze long records of natural phenomena \cite{eke02,hurst65}.
There are two factors used in this analysis: firstly the range $R$,
this is the difference between the minimum and maximum 'accumulated'
values or cumulative sum of $X(t,\tau)$ of a typical natural
phenomenon at discrete time $t$ over a time span $\tau$, and
secondly the standard deviation $S$, estimated from the observed
values $X(t,\tau)$. Hurst found that the ratio $R/S$ is very well
described for a large number of natural phenomena by the following
empirical relation
\begin{equation} R/S \sim \tau^H,\label{RS1} \end{equation}
where $\tau$ and $H$ are time span and Hurst exponent,
respectively. For temperature fluctuations on the last scattering surface,
$R$ and $S$ are defined as
\begin{eqnarray} R(\gamma_s)&=& {\rm Max}\{Y(\gamma_s)\}-{\rm
Min}\{Y(\gamma_s)\},\\
S(\gamma_s)&=& \left(\frac{1}{s}\sum_{i=1}^{s} \left[ T(\hat{n}_i)
- \langle T \rangle \right]^2\right)^{1/2},\hskip 0.1cm
s=1,\ldots,N,\end{eqnarray} where $Y(\gamma_s)$ is defined
according to Eq. (\ref{profile}). The scaling behavior of the
fluctuation function is determined by analyzing log-log plot of
$R/S$ versus $\gamma_s$ as
\begin{equation}
R/S \sim \gamma_s^H.
\end{equation}

\section{Fractal Analysis of cosmic microwave background radiation data}

\begin{figure}
\epsfxsize=8truecm\epsfbox{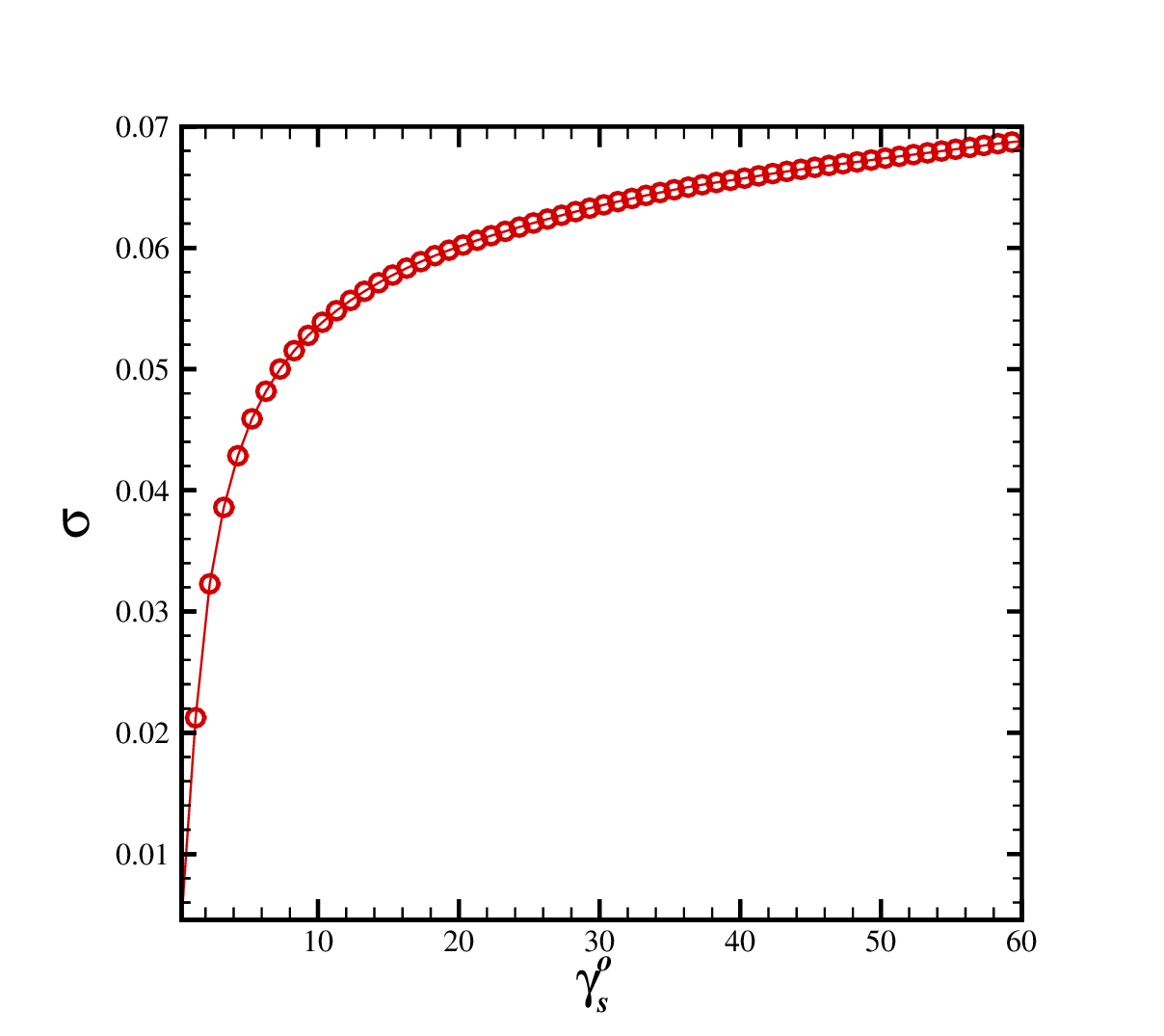} \narrowtext \caption{ (Color online)
Behavior of standard deviation of temperature fluctuations of CMB
as a function of angular scale, $\gamma_s$.} \label{fig12}
 \end{figure}

As mentioned in section II, a spurious of correlations may be
detected in the absence of statistical isotropy, so direct
calculation of correlation functions, fractal dimensions etc., may
not give reliable results \cite{Peng94,taqqu,feder88,eke02,bun02}.
The simplest way to verify the statistical isotropy of the
temperature fluctuations on the last scattering surface is by
measuring the stability of the variance of the temperature in
various sizes of windows. Figure~\ref{fig12} shows the standard
deviation of the temperature verses the angular size $\gamma_s$ of
the window. Here we have a saturation for the standard derivation in
the large angular scale which reflects the long-range correlation
behavior of the temperature fluctuations of CMB
\cite{Peng94,taqqu,feder88,eke02,bun02}.

\begin{figure}
\epsfxsize=8truecm\epsfbox{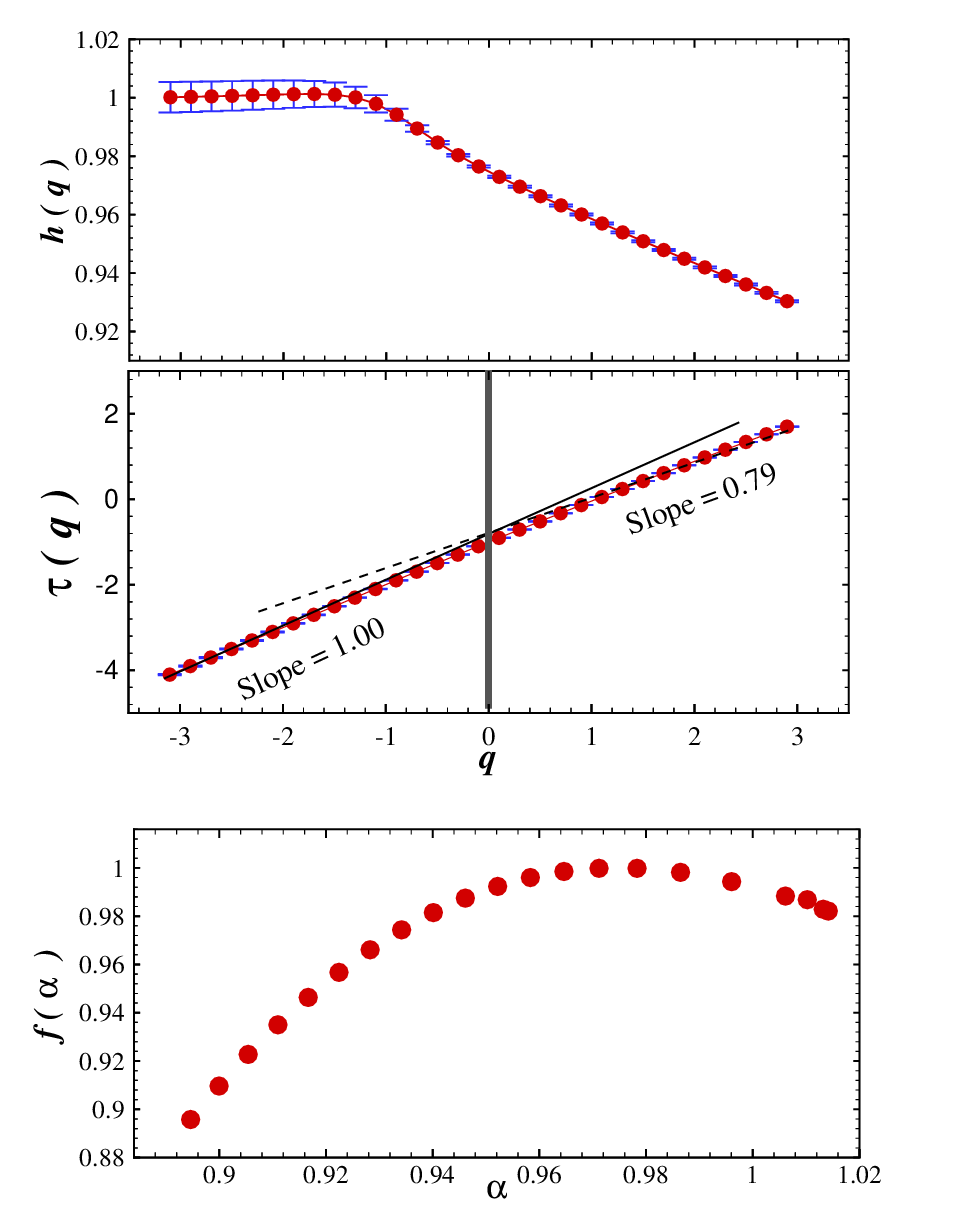} \narrowtext \caption{ (Color online) The $q$
dependence of the generalized Hurst exponent $h(q)$, the
corresponding $\tau(q)$ and singularity spectrum $f(\alpha)$ are
shown in the upper to lower panel respectively for temperature
fluctuation series.} \label{fig2}
 \end{figure}

We use the MF-DFA1 method for analyzing the temperature fluctuations
on CMB. Following the procedure for MF-DFA analysis as described in
the last section we obtain $F_q(\gamma_s)$ as a function of angular
scale $\gamma_s$. For each index of $q$ the generalized Hurst
exponents $h(q)$ in Eq.~(\ref{Hq}) can be found by analyzing log-log
plots of $F_q(\gamma_s)$ versus $\gamma_s$. Figure \ref{fig2} shows
the Hurst exponent in terms of $q$ for MF-DFA1 analysis. Variation
of Hurst exponent in terms of $q$ shows the multifractal behavior of
temperature fluctuation of CMB. More over we have different values
of the slope of classical multifractal scaling exponent,$\tau(q)$,
for $q<0$ and $q>0$, which indicates that CMB has a multifractal
structure. For positive and negative values of $q$, we obtain
$\tau(q)$ with the slopes of $0.79\pm0.03$ and $1.00\pm0.03$,
respectively  at $1\sigma$ confidence interval. According to the
relation between the Hurst exponent and MF-DFA exponent, we obtain
$H=0.94\pm0.01$ at $1\sigma$ confidence interval which means that
temperature fluctuation is an isotropic process with long-range
correlation \cite{bun02}. The variation of singularity spectrum
$f(\alpha)$, Eq. (\ref{Legendre}) also is shown in Figure
\ref{fig2}. The values of derived quantities from MF-DFA1 method,
are given in Table \ref{Tab1}.

\begin{figure}
\epsfxsize=8truecm\epsfbox{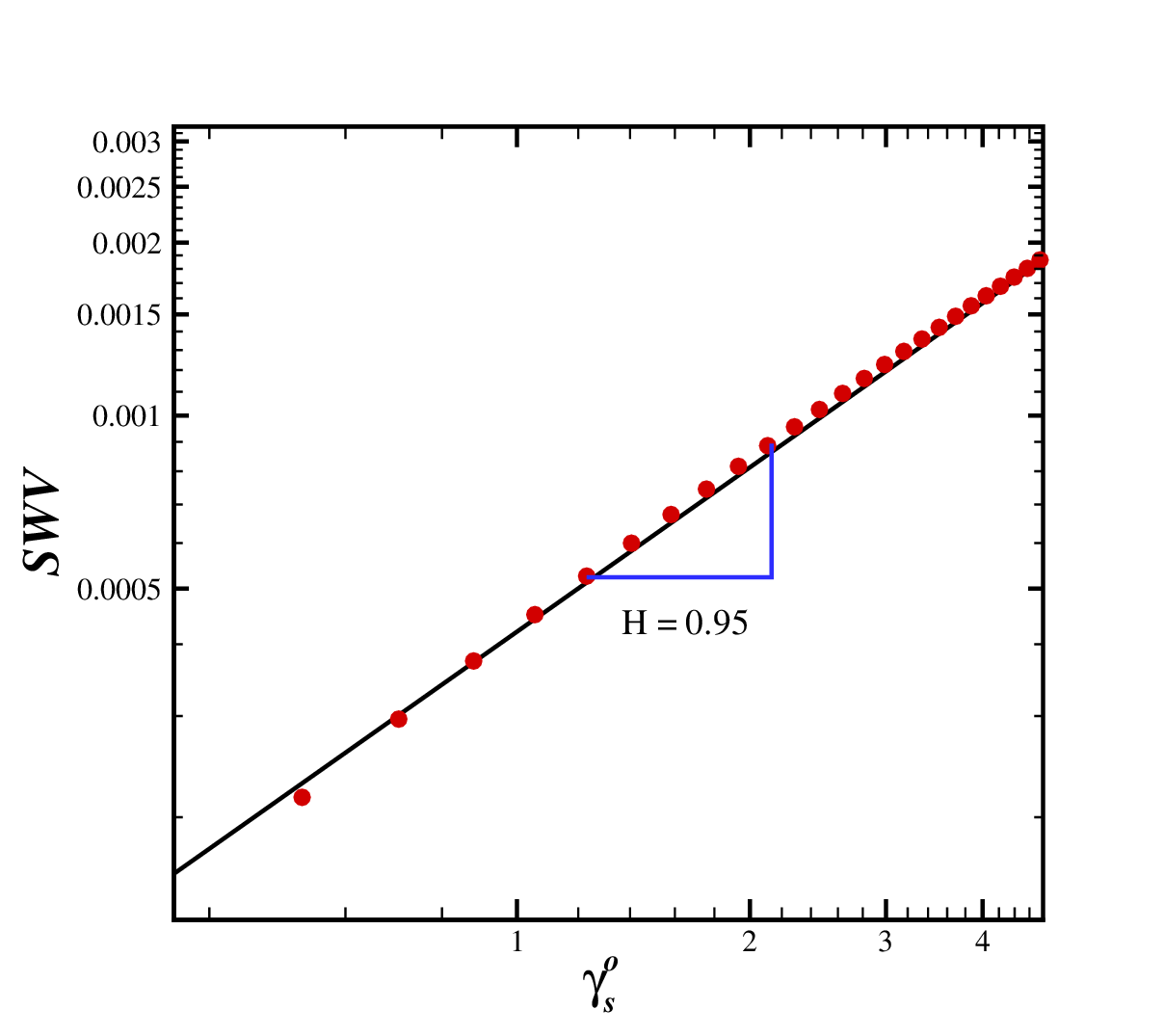} \narrowtext \caption{(Color online) Behavior
of Scaled Windowed Variance (SWV) of CMB fluctuations as a function
of angular scale, $\gamma_s$ in log-log plot.} \label{fig_swv}
 \end{figure}

\begin{figure}
\epsfxsize=8truecm\epsfbox{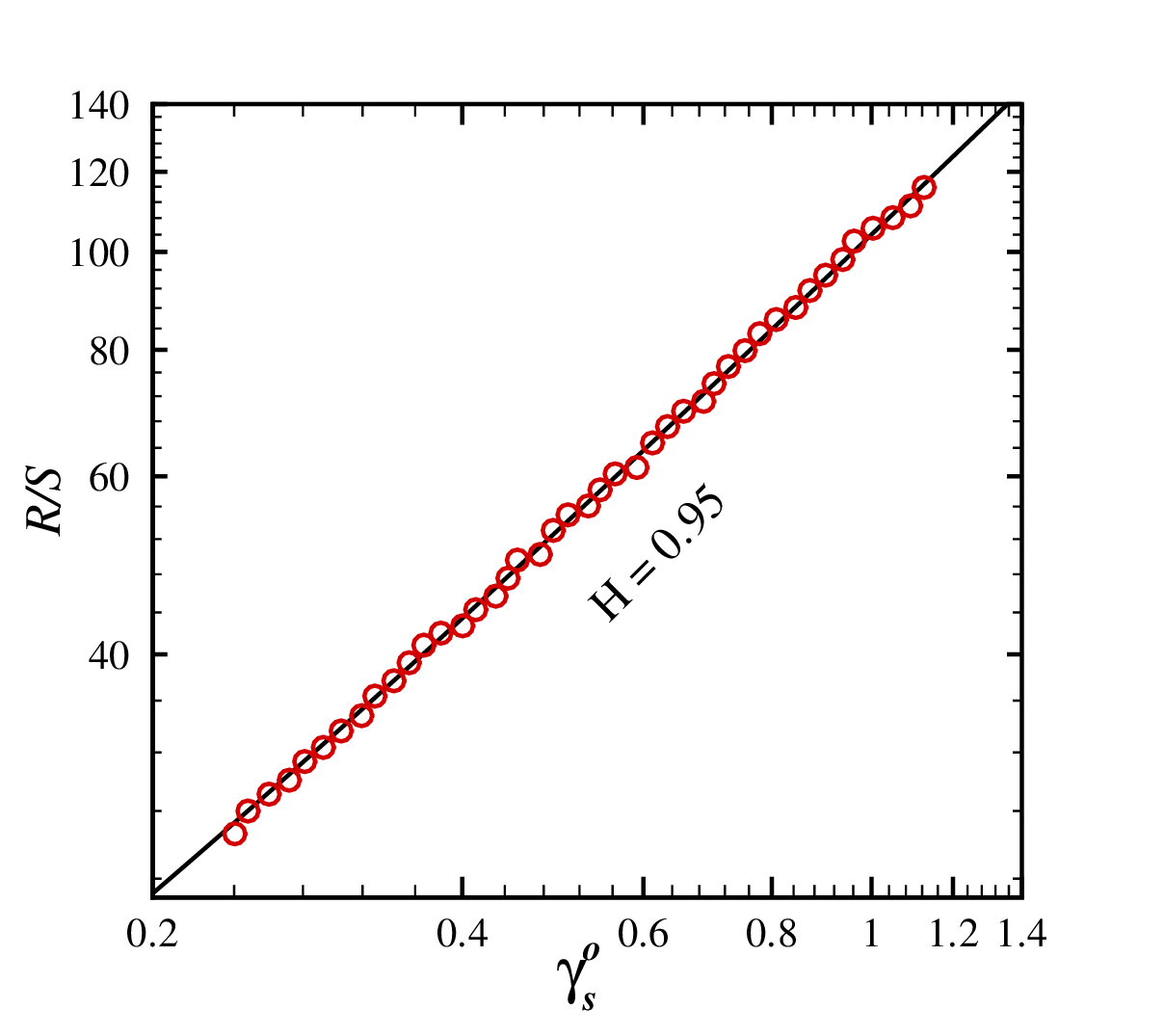} \narrowtext \caption{(Color online) The
log-log plot of Rescaled Range ($R/S$) versus angular scale,
$\gamma_s$ for temperature fluctuation at last scattering surface.}
\label{fig_rs}
 \end{figure}

We also use the Scaled Windowed Variance (SWV) analysis
\cite{eke02,hurst65} to determine the Hurst exponent for CMB via Eq.
(\ref{swv}). Figure \ref{fig_swv} shows log-log plot of Scaled
Windowed Variance of CMB fluctuations as a function of angular
scale, $\gamma_s$ which results $H=0.95\pm0.02$ at $68\%$ confidence
interval. Finally we apply the Rescaled Range analysis ($R/S$), to
determine the Hurst exponent of CMB fluctuations. According to
Figure \ref{fig_rs}, the value of Hurst exponent obtain as
$0.95\pm0.02$ at $68\%$ confidence interval, which is in agreement
with the two previous results.

%%%%%%%%%%%%%%%%%%%%%%%%%%%%%%%%%%%%%%%%%%%%%%%%%%%%%%%%%%%%%%%%%%%%%%%%%
%%%%%%%%%%%%%%%%%%%%%%%%%%%%%%%%%%%%%%%%%%%%%%%%%%%%%%%%%%%%%%%%%

%%%%%%%%%%%%%%%%%%%%%%%%%%%%%%%%%%%%%%%%%%%%%%%%%%%%%%%%%%%%%%%%%%%%%%%%%
%%%%%%%%%%%%%%%%%%%%%%%%%%%%%%%%%%%%%%%%%%%%%%%%%%%%%%%%%%%%%%%%%
%%%%%%%%%%%%%%%%%%%%%%%%%%%%%%%%%%%%%%%%%%%%%%%%%%%%%%%%%%%%%%%%%%%%%%%%%
%%%%%%%%%%%%%%%%%%%%%%%%%%%%%%%%%%%%%%%%%%%%%%%%%%%%%%%%%%%%%%%%%
%%%%%%%%%%%%%%%%%%%%%%%%%%%%%%%%%%%%%%%%%%%%%%%%%%%%%%%%%%%%%%%%%%%%%%%%%
%%%%%%%%%%%%%%%%%%%%%%%%%%%%%%%%%%%%%%%%%%%%%%%%%%%%%%%%%%%%%%%%%

\section{Comparison of  multifractal nature for original,
shuffled and surrogate CMB data}

In this section we are interested in to determine the source of
multifractality and test the Gaussianity of CMB data. In general,
two different types of multifractality in certain data set can be
distinguished: (i) Multifractality due to a fatness of probability
density function (PDF) and (ii) Multifractality due to different
correlations in small and large scale fluctuations. In the first
case the multifractality cannot be removed by shuffling the data
set, while in the second case data may have a PDF with finite
moments, e.~g. a Gaussian distribution and the corresponding
shuffled series will exhibit mono-fractal scaling, since all
long-range correlations are destroyed by the shuffling procedure.

\begin{table}[htb]
\caption{\label{Tab1}The values of the Hurst, multifractal scaling
 and generalized multifractal exponents for $q=2.0$ at $1\sigma$ confidence interval,
for original, surrogate and shuffled of temperature fluctuation
series obtained by MF-DFA1.}
%\medskip
\begin{center}
\begin{tabular}{|c|c|c|c|}
%\hline
    Data & $H$ & $\tau$ &$D$ \\ \hline
   CMB  & $0.94\pm 0.01$ &$0.88\pm0.02$  &$0.88\pm0.02$     \\\hline
   Surrogate &$0.88\pm0.01$ &$0.76\pm0.02$ &$0.76 \pm 0.02$     \\\hline
  Shuffled & $0.500\pm0.001$ &$0.002\pm0.002$  & $0.002\pm0.002$  \\ %\hline
\end{tabular}
\end{center}
\end{table}

If we have both kinds of multifractality in data, the shuffled
series will show weaker multifractality than the original series.
The easiest way to distinguish the type of multifractality, is by
analyzing the corresponding shuffled and surrogate series. The
shuffling of data set destroys the long range correlation, therefore
if the multifractality only belongs to the long-range correlation,
we should find $h_{\rm shuf}(q) = 0.5$ \cite{bun02}. The
multifractality nature due to the fatness of the PDF signals is not
affected by the shuffling procedure. On the other hand, to determine
the multifractality due to the broadness of PDF by the surrogating
method, the phase of discrete fourier transform (DFT) coefficients
of CMB data are replaced with a set of pseudo independent uniform
distribution within the range of $(-\pi,\pi)$ \cite{sa05b,pric94}.
The correlations in the surrogate series do not change, while the
probability function changes to the Gaussian distribution. If
multifractality in the data set is due to a broadness of PDF, $h(q)$
obtained by the surrogate method  will be independent of $q$. If
both kinds of multifractality are present in CMB data, the shuffled
and surrogate series will show weaker multifractality than the
original one.

 To check the nature of multifractality, we compare the
fluctuation function $F_q(\gamma_s)$, for the CMB map with the
result of the corresponding shuffled, $F_q^{\rm shuf}(\gamma_s)$ and
surrogated data $F_q^{\rm sur}(\gamma_s)$. Differences between these
two fluctuation functions with the original one can indicate the
presence of long range correlations or broadness of probability
density function in the original data. These differences can be
obtained by the ratio $F_q(\gamma_s) / F_q^{\rm shuf}(\gamma_s)$ and
$F_q(\gamma_s) / F_q^{\rm sur}(\gamma_s)$ as a function of
$\gamma_s$ \cite{bun02}. Since the anomalous scaling due to a broad
probability density affects $F_q(\gamma_s)$ and $F_q^{\rm
shuf}(\gamma_s)$ in the same way, only multifractality due to
correlations will be observed in $F_q(\gamma_s) / F_q^{\rm
shuf}(\gamma_s)$. The scaling behavior of these ratios are:
\begin{equation} F_q(\gamma_s) / F_q^{\rm shuf}(\gamma_s) \sim \gamma_s^{h(q)-h_{\rm
shuf}(q)} = \gamma_s^{h_{\rm corr}(q)}. \label{HqCor}
\end{equation}
\begin{equation} F_q(\gamma_s) / F_q^{\rm sur}(\gamma_s) \sim \gamma_s^{h(q)-h_{\rm
sur}(q)} = \gamma_s^{h_{\rm PDF}(q)}. \label{Hqpdf} \end{equation}
If only fatness of the PDF be responsible for the multifractality,
one should find $h(q)=h_{\rm shuf}(q)$ and $h_{\rm corr}(q)=0$. On
the other hand, deviations from $h_{\rm corr}(q) =0$ indicates the
presence of correlations, and $q$ dependence of $h_{\rm corr}(q)$
indicates that multifractality is due to the long-rage
correlation.

\begin{figure}
\epsfxsize=8truecm\epsfbox{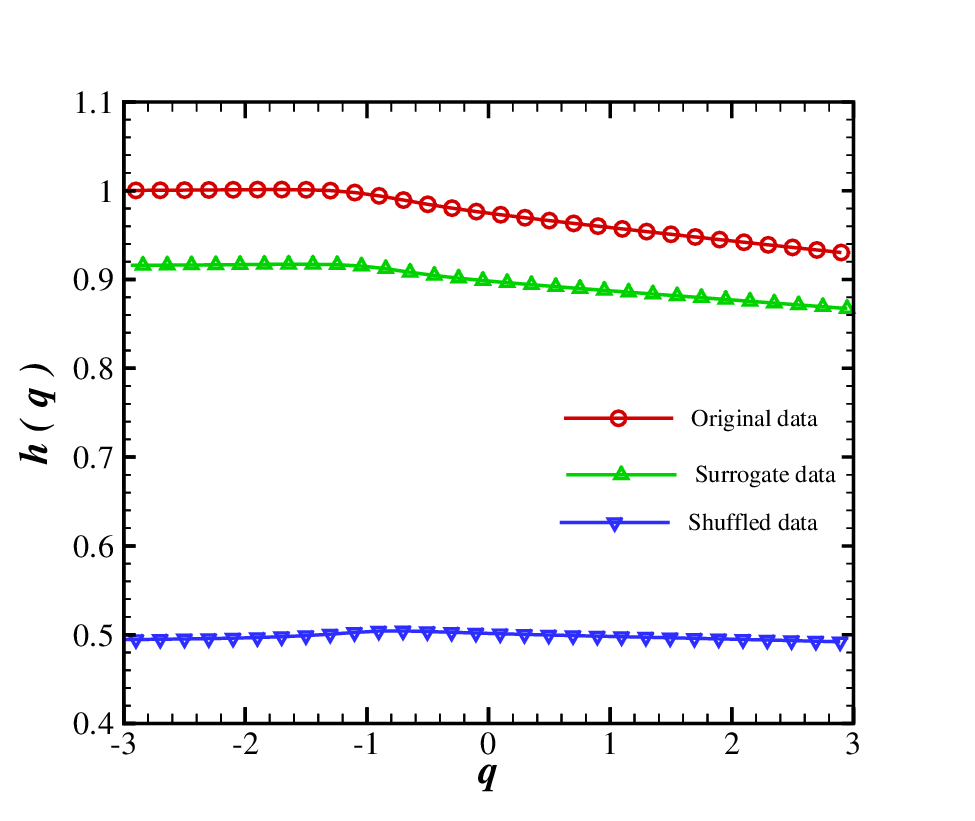} \narrowtext
\caption{ (Color online) Generalized Hurst exponent as a function of $q$ for
 original, surrogate and shuffled CMB data.}
\label{fig5}
 \end{figure}

If the multifractal behavior of CMB is made by the broadness of
PDF and long-range correlation, both $h_{\rm shuf}(q)$ and $h_{\rm
sur}(q)$ will depend on $q$. The $q$ dependence of the generalized
exponent $h(q)$ for original, surrogate and shuffled CMB data are
shown in Figures \ref{fig5} which shows that the multifractality
 nature of temperature fluctuation  is almost due
 to the $long-range$ correlation. The absolute value of $h_{\rm corr}(q)$
 is larger than $h_{\rm PDF}(q)$ which indicates that the multifractality due
  to the fatness is much weaker than the multifractality
 due to the correlation. The deviation of $h_{\rm sur}(q)$ and $h_{\rm shuf}(q)$
 from $h(q)$ can be determined by using $\chi^2$ as follows:

\begin{equation}
 \chi^2_{\diamond}=\sum_{i=1}^{N}\frac{[h(q_i)-h_{\diamond}(q_i)]^2}{\sigma(q_i)^2+\sigma_{\diamond}(q_i)^2},
 \label{khi} \end{equation}

where the symbol $"\diamond"$ represents for $"\rm surroated"$ and
$"\rm shuffled"$ series. The value of reduced chi-square
$\chi^2_{\nu\diamond}=\chi^2_{\diamond}/\cal{N}$ ($\cal{N}$ is the
number of degree of freedom) for shuffled and surrogate time series
are  $21599$, $313$, respectively.

\begin{figure}
\epsfxsize=8truecm\epsfbox{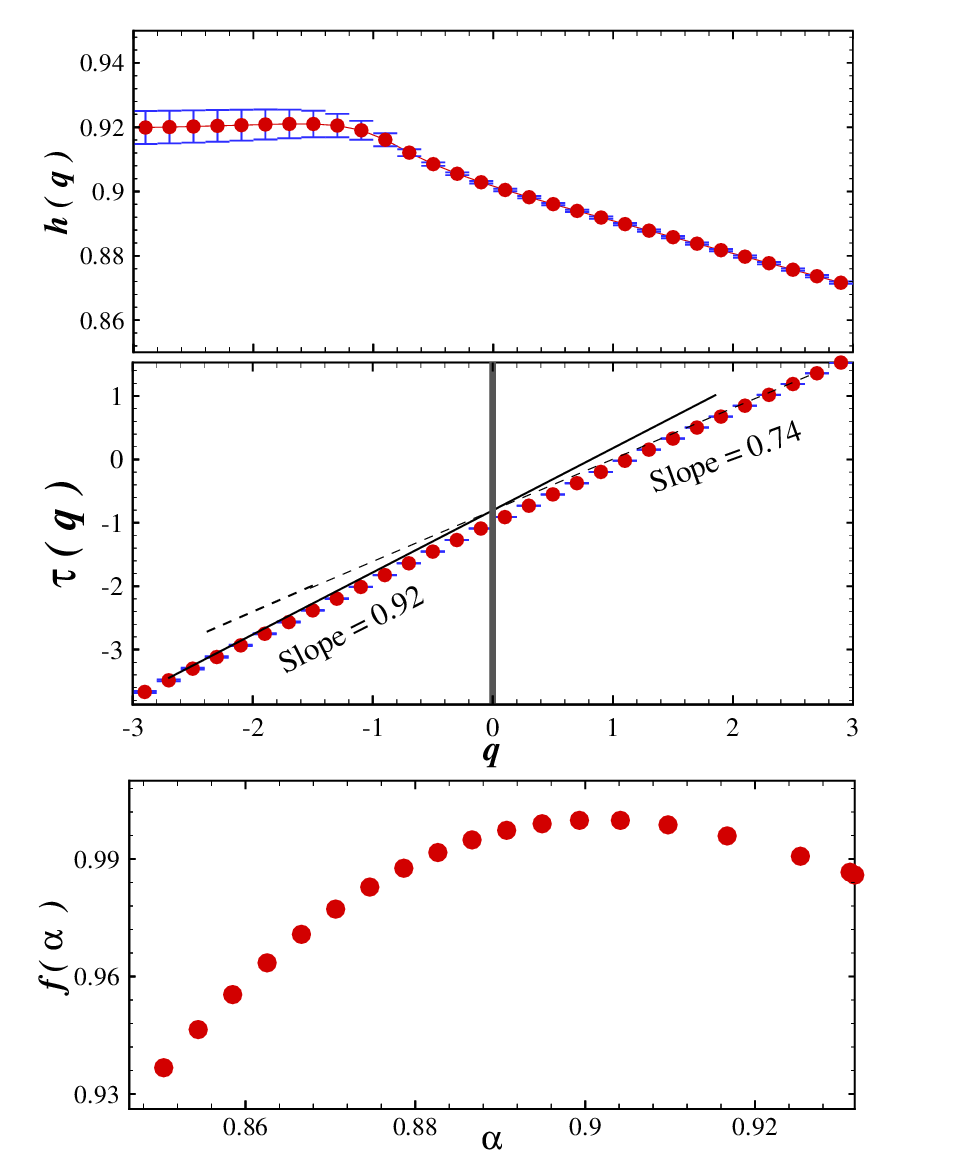} \narrowtext \caption{ (Color online) The $q$
dependence of the generalized Hurst exponent $h(q)$, the
corresponding $\tau(q)$ and singularity spectrum $f(\alpha)$ are
shown in the upper to lower panel respectively for surrogate
temperature fluctuation series.} \label{fig3}
 \end{figure}

\begin{figure}
\epsfxsize=8truecm\epsfbox{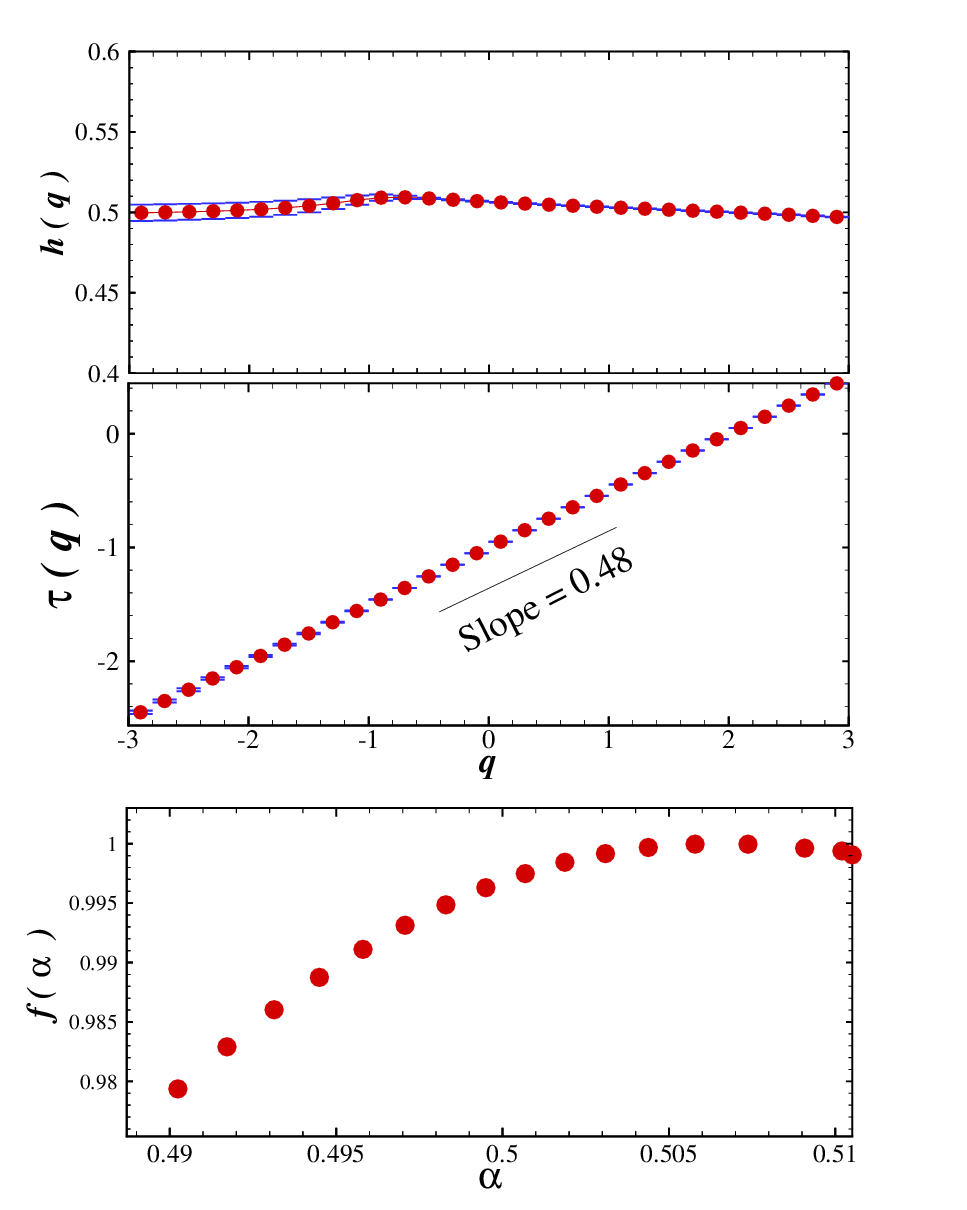} \narrowtext \caption{ (Color online) The $q$
dependence of the generalized Hurst exponent $h(q)$, the
corresponding $\tau(q)$ and singularity spectrum $f(\alpha)$ are
shown in the upper to lower panel respectively for shuffled
temperature fluctuation series.} \label{fig4}
 \end{figure}

Singularity spectrum $f(\alpha)$ of the surrogate series is almost
similar to the original temperature fluctuations, while in the case
of shuffled series we have a narrower singularity spectrum of
$\Delta\alpha_{\rm shuf}=\alpha(q_{min})-\alpha(q_{max})\simeq0.02$
compare to that of surrogate  $\Delta\alpha_{\rm sur}\simeq0.09$ and
original data $\Delta\alpha \simeq0.12$. The small value of
$\Delta\alpha_{\rm shuf}$ compare to the two other series shows that
the multifractality in the shuffled CMB map has almost been
destroyed \cite{paw05}. Figures \ref{fig3} and \ref{fig4} show the
MF-DFA1 results for surrogate and shuffled temperature fluctuation
series. Comparing the MF-DFA results of the data set to those for
shuffled and surrogate series, we conclude that the multifractal
nature of temperature fluctuations in CMB data is almost due to the
$long-range$ $correlations$ and the probability distribution
function of CMB is almost Gaussian (see also the section $V$). The
values of the Hurst exponent $H$, multifractal scaling $\tau(q=2)$,
generalized multifractal dimension $D(q=2)$  of the original,
shuffled and surrogate data of CMB obtained with MF-DFA1 method are
reported in Table \ref{Tab1}.

%%%%%%%%%%%%%%%%%%%%%%%%%%%%%%%%%%%%%%%%%%%%%%%%%%%%%%%%%%%%%%%%%%%%%%%%%
%%%%%%%%%%%%%%%%%%%%%%%%%%%%%%%%%%%%%%%%%%%%%%%%%%%%%%%%%%%%%%%%%

%%%%%%%%%%%%%%%%%%%%%%%%%%%%%%%%%%%%%%%%%%%%%%%%%%%%%%%%%%%%%%%%%%%%%%%%%
%%%%%%%%%%%%%%%%%%%%%%%%%%%%%%%%%%%%%%%%%%%%%%%%%%%%%%%%%%%%%%%%%

%%%%%%%%%%%%%%%%%%%%%%%%%%%%%%%%%%%%%%%%%%%%%%%%%%%%%%%%%%%%%%%%%%%%%%%%%
%%%%%%%%%%%%%%%%%%%%%%%%%%%%%%%%%%%%%%%%%%%%%%%%%%%%%%%%%%%%

\section{Statistical Isotropy and Gaussianity of CMB Anisotropy}

In this section we examine the Gaussianity and statistical isotropy
of temperature fluctuations in CMB data through MF-DFA, $R/S$ and
SWV methods.

 Standard models of inflationary cosmology predict
a Gaussian distribution of CMB temperature fluctuations. However,
along with standard inflationary models, there exist theories such
as inflationary scenarios with two or more scalar fields which
predict a non-Gaussian primordial fluctuations
\cite{linde97,anto97,pee99a,pee99b,movahedstring,starck,firouz05}.
Another possibility of deviation from the Gaussianity is the
manipulation of the CMB signals after the recombination due to
subsequent weak gravitational lensing \cite{fuk95,ber97} and various
foregrounds like dust emission, synchrotron radiation, or unresolved
point sources \cite{ban96}. One should also take into account the
additional instrumental noise in the observational data
\cite{teg97}. One of the advantages of Gaussian field is that the
two point correlation function
$C(\hat{n},\,\hat{n}^\prime)\equiv\langle T(\hat n) T(\hat
n^\prime)\rangle $ will be sufficient for fully specify the
statistical properties of the field. In the case of non-Gaussianity,
one must take the higher moments of the field into account in order
to fully describe the whole statistics. So, studying the Gaussian
nature of the signals or detecting some distinctive non-Gaussianity
is an important issue to understand the statistical properties of
CMB.

As we have seen in  Sec.IV, the MF-DFA1 results of the shuffled
series shows almost constant $h(q)\simeq 0.5$ which indicates that
$h(q)$ is independent of the $q$ or in another word the shuffled
series is a monofractal structure. This means that the
multifractality of the original data is due to the long-range
correlations not due to a broadness of PDF of temperature
fluctuations (Figure \ref{fig5}). Furthermore small differences in
the generalized Hurst exponent between the original and surrogated
data (Figure \ref{fig5}) and similarity of the singularity spectrum,
$f(\alpha)$, in those two series, show that the probability density
function of temperature fluctuations has well confidence with the
Gaussian distribution which is in agreement with the recent results
\cite{hajian05,hajian04,sree03,sa05a}

The second important question about the CMB data is its statistical
isotropy. One of the main motivations for anisotropy of the universe
comes from the global topology of the universe. The idea of a
multi-connected topology of the universe comes back to
Schwarzschild's work \cite{sch1900} before the Einstein's first
seminal paper on cosmology \cite{ein1917}. If the universe is
assumed to be simply-connected, the cosmological principle implies
that it should be globally isotropic to any observer. In the case of
multi-connected universe while the Einstein gravity still hold in
this space but the global structure of universe at large scales will
be more complicated. One of the most promising signatures may be on
the CMB as the existence of large scale anisotropies with a
repeating patterns \cite{top}. In addition to cosmological
mechanisms, instrumental and environmental effects in observations
of CMB can also easily produce spurious breakdown of statistical
isotropy. Here we examine the anisotropy of CMB with the MF-DFA
analysis.

 The statistical isotropy means that the
statistical properties of CMB (e.g. $n$-point correlations) to be
invariant under the Eulerian rotation. This means that two point
correlation function is function of angle between two points, i.e.
\begin{equation}
C(\hat{n}_1,\hat{n}_2) = C(\hat{n}_1 \cdot \hat{n}_2)\equiv
C(\gamma),
\end{equation}
 where $\gamma=\arccos(\hat{n}_1 \cdot
\hat{n}_2)$ is the angle between $\hat{n}_1$ and $\hat{n}_2$. It
is then convenient to expand it in terms of Legendre polynomials,
\begin{equation} C(\gamma) \,=\,
\frac{1}{4\pi}\sum_{l=2}^{\infty}(2l+1) C_l P_l(\cos{\gamma}),
\end{equation}
 where $C_l$ is the widely used {\it angular power spectrum}. The summation over $l$ starts from
$2$ because the $l\,=\,0$ term is the monopole which in the case of
statistical isotropy the monopole is constant, and it can be
subtracted out. The dipole $l\,=\,1$ is due to the local motion of
the observer with respect to the last scattering surface and can be
subtracted out as well. In order to extract the statistical property
of CMB we need an ensemble maps from the CMB to calculate the
correlation functions. In reality we have only one map from the CMB,
but if we assume the statistical isotropy we can calculate the
correlation functions moving on overall data. As we mentioned in
section II, for a process with stationarity in time or isotropic
process in space, the Hurst exponent is always less than unity
\cite{Peng94,eke02}. According to the MF-DAF1, SWV and $R/S$ methods
we obtained the value of Hurst exponent for {\it WMAP} data as $H=
0.94\pm0.02$ at $1\sigma$ confidence interval. These analysis with
the Hurst exponent $(0<H<1.0)$ show that there is no evidence for
violation of statistical isotropy of {\it WMAP} data.

\section{Conclusion}

In this work we used three method to investigate the statistical
properties of {\it WMAP} data. The applied methods
 Multifractal
Detrended Fluctuation Analysis, Rescaled Range Analysis and Scaled
Windowed Variance methods show that the {\it WMAP} data is a
statistically isotropic data set. The value of Hurst exponent,
$H\simeq0.94$, guarantees that there is no evidence for violation of
statistical isotropy in the CMB map. The MF-DFA method allows us to
determine the multifractal characterization of this map. The $q$
dependence of generalized Hurst $h(q)$ and classical multifractal
scaling $\tau(q)$ exponents, show that the temperature fluctuations
on the last scattering surface has multifractal behavior
\cite{sree03}. We obtained the generalized multifractal dimension of
CMB data $D\simeq0.88$. According to a small difference between
generalized Hurst exponent and singularity spectrum of the original
data set with the surrogate one's, $H-H_{{\rm sur}}=0.06\pm0.01$ and
$\Delta\alpha-\Delta\alpha_{{\rm sur}}=0.03$, we found that there is
no crucial evidence for non-Gaussianity of the CMB map. The compatibility of the
multifractal behavior of the
temperature fluctuations of CMB with their Gaussian distribution has
been also explored recently \cite{Ber}. 
Comparison
of the generalized Hurst exponent of the original data with the
shuffled and surrogate series showed that the nature of the
multifractality of CMB is mainly due to the long-range correlation ,
rather than the broadness of the probability density function.

{\bf Acknowledgements}

We would like to thank M. Fazeli and G.R. Jafari for useful discussions.

\end{document}